\documentclass[12pt,preprint]{aastex}

\newcommand{\teff}{$T_{\rm eff}$}
\newcommand{\tc}{$T_{\mathrm{c}}$}
\newcommand{\cyga}{16\,Cyg\,A}
\newcommand{\cygb}{16\,Cyg\,B}
\newcommand{\cygab}{16\,Cyg\,A and B}

\begin{document}

\title{DETAILED ABUNDANCES OF THE SOLAR TWINS 16 CYGNI A AND B:  CONSTRAINING 
PLANET FORMATION MODELS\altaffilmark{*}}

\altaffiltext{*}{The data presented herein were obtained at the W.M. Keck 
Observatory, which is operated as a scientific partnership among the California 
Institute of Technology, the University of California, and the National 
Aeronautics and Space Administration. The Observatory was made possible by the 
generous financial support of the W.M. Keck Foundation.}

\author{
Simon C. Schuler\altaffilmark{1,2}, Katia Cunha\altaffilmark{1,3,4}, Verne V. 
Smith\altaffilmark{1}, Luan Ghezzi\altaffilmark{3,5}, Jeremy R. 
King\altaffilmark{6}, Constantine P. Deliyannis\altaffilmark{7}, AND Ann Merchant
Boesgaard\altaffilmark{8}
	}
\affil{
\altaffiltext{1}{National Optical Astronomy Observatory, 950 North Cherry 
Avenue, Tucson, AZ, 85719  USA; sschuler@noao.edu, kcunha@noao.edu, 
vsmith@noao.edu}
\altaffiltext{2}{Leo Goldberg Fellow}
\altaffiltext{3}{Observat{\'o}rio Nacional, Rua General Jos{\'e} Cristino, 77, 
20921-400, S{\~a}o Crist{\'o}v{\~a}o, Rio de Janeiro, RJ, Brazil; luan@on.br}
\altaffiltext{4}{Steward Observatory, University of Arizona, 933 North Cherry 
Avenue, Tucson, AZ 85721  USA}
\altaffiltext{5}{Laborat{\'o}rio Interinstitucional de e-Astronomia, - LIneA, 
Rua Gal. Jos{\'e} Cristino 77, Rio de Janeiro, RJ - 20921-400, Brazil}
\altaffiltext{6}{Department of Physics and Astronomy, Clemson University, 118 
Kinard Laboratory, Clemson, SC, 29634  USA; jking2@ces.clemson.edu}
\altaffiltext{7}{Department of Astronomy, Indiana University, Swain Hall West 
319, 727 East 3rd Street, Bloomington, IN 47405-7105  USA; con@astro.indiana.edu}
\altaffiltext{8}{Institute for Astronomy, University of Hawaii at Manoa, 2680 
Woodlawn Drive, Honolulu, HI 96822, USA;  boes@ifa.hawaii.edu}
}

\begin{abstract}
Results of a detailed abundance analysis of the solar twins \cyga\ and \cygb\ 
based on high-resolution, high signal-to-noise ratio echelle spectroscopy are 
presented.  \cygb\ is known to host a giant planet while no planets have yet been 
detected around \cyga.  Stellar parameters are derived directly from our 
high-quality spectra, and the stars are found to be physically similar, with 
$\Delta$\teff\ $=+43$ K, $\Delta \log g=-0.02$ dex, and $\Delta \xi=+0.10$ km 
s$^{-1}$ (in the sense of A $-$ B), consistent with previous findings. 
Abundances of 15 elements are derived and are found to be indistinguishable 
between the two stars.  The abundances of each element differ by $\leq 0.026$ 
dex, and the mean difference is $+0.003 \pm 0.015$ ($\sigma$) dex.  Aside from Li, 
which has been previously shown to be depleted by a factor of at least 4.5 in 
\cygb\ relative to \cyga, the two stars appear to be chemically identical.  The 
abundances of each star demonstrate a positive correlation with the condensation 
temperature of the elements (\tc); the slopes of the trends are also 
indistinguishable.  In accordance with recent suggestions, the positive 
slopes of the [m/H]-\tc\ relations may imply that terrestrial planets have not 
formed around either \cyga\ or \cygb.  The physical characteristics of the 16 Cyg 
system are discussed in terms of planet formation models, and plausible 
mechanisms that can account for the lack of detected planets around \cyga, the 
disparate Li abundances of \cygab, and the eccentricity of the planet \cygb\ b 
are suggested.
\end{abstract}

\keywords{planetary systems:formation -- stars:abundances -- stars:atmospheres --
stars:individual(16 Cyg A, 16 Cyg B)}

\section{INTRODUCTION}
\label{s:intro}
\cyga\ and \cygb\ are a well known common proper-motion pair of solar-twin stars 
with spectral types G1.5V and G3V, respectively.  Stellar parameters and 
[Fe/H] abundances of the pair have been derived by numerous groups
\citep[e.g.,][]{1994PASP..106.1248G,1996A&A...311..245F,1998A&A...336..942F,2001ApJ...553..405L,2005PASJ...57...83T},
and the abundances of additional elements have been derived by others 
\citep[e.g.,][]{1993A&A...274..825F,1997AJ....113.1871K,1998A&AS..129..237F,1998A&A...334..221G,2000AJ....119.2437D,2001PASJ...53.1211T,2003MNRAS.340..304R,2004ARep...48..492G}.
In each study, \cygab\ have been found to be physically similar, with A being
slightly hotter and having a slightly lower surface gravity than B, consistent
with their spectral types.  Differences in the derived stellar parameters in
the sources listed above range from +25 to +62 K in \teff, -0.03 to -0.15 dex in 
$\log g$, and 0 to +0.05 dex in [Fe/H] (all comparisons herein are made in the 
sense of A $-$ B).

A defining property distinguishing the two stars is the designation of \cygb\ as 
a planet host.  \citet{1997ApJ...483..457C} reported the presence of a 
radial-velocity detected planet (\cygb\ b) with $M \sin i=1.5 \; 
M_{\mathrm{Jup}}$ orbiting \cygb\ on an eccentric orbit ($e=0.63$), but 
despite being monitored with the same temporal coverage, no planet was detected 
around \cyga.  Continued radial-velocity monitoring has yielded no additional 
planet signatures for either star (D. Fischer, private communication).  Imaging 
observations, however, do indicate that \cyga\ has a faint M dwarf binary 
companion with a separation of $\sim 3$'', corresponding to projected separation 
of $\sim 70$ AU at the measured distance of the system
\citep[$\sim 22$ pc;][]{1999PASP..111..321H,2001AJ....121.3254T,2002ApJ...581..654P}.  
Whether these two objects are gravitationally bound has yet to be determined 
firmly, but initial proper motion measurements do suggest that they are 
physically associated \citep{2002ApJ...581..654P}.

\citet{1993A&A...274..825F} and subsequently \citet{1997AJ....113.1871K} found 
that \cygab\ differ in another fundamental way: their Li abundances.  The 
photospheric Li abundance of \cygb\ is a factor $\geq 4.5$ lower than that of 
\cyga.  While both stars are depleted in Li relative to the Solar System's 
meteoritic value \citep[$\log N(\mathrm{Li})=3.26$;][]{2009ARA&A..47..481A}, 
the Li abundance of \cyga\  ($\log N(\mathrm{Li})=1.27$) is slightly higher 
and that of \cygb\ is lower ($\log N(\mathrm{Li}) \leq 0.60$) than that of the 
Sun ($\log N_{\odot}(\mathrm{Li})=1.05$) \citep{1997AJ....113.1871K}.  The 
difference in the Li abundances of \cygab\ cannot be explained by standard 
stellar models, which predict Li depletion is a function of stellar age, mass, 
and composition; empirical evidence suggests that an extra parameter is needed.  
\citet{1997AJ....113.1871K} argue that a slow mixing mechanism, possibly related 
to rotation, can account for the low absolute Li abundances of both stars, and 
they discuss a possible connection between Li depletion and planet formation as an 
explanation for the difference between the two.  More recently, others have also
argued that an extra parameter (beyond standard models) is needed to account for 
the observed Li abundances of solar-type stars 
\citep[e.g.,][]{2008A&A...489..677P}.  \citet{2000AJ....119.2437D} note that the 
Li--\teff\ trend could be quite steep for solar twins, consistent with the 16 Cyg 
A -- Sun -- 16 Cyg B pattern, so that even if initial angular momentum 
($J_{\mathrm{o}}$) and rotational history do play the role of the extra parameter, 
$J_{\mathrm{o}}$ need not be unreasonably different between A and B.  
\citet{2000AJ....119.2437D} also found that the Be abundances of \cygab\ are the 
same within the measurement uncertainties, placing an additional constraint on the 
mechanism responsible for the disparate Li abundances.

In this Letter we present the results of a detailed abundance analysis of 15
elements of the solar twins \cyga\ (HR\,7503, HD\,186408, HIP\,96895) and 
\cygb\ (HR\,7504, HD\,186427, HIP\,96901) based on high-resolution echelle 
spectroscopy.  The abundances allow us to constrain more fully the physical 
similarities of the two stars, and the implications for Li depletion and planet
formation in this system are discussed.

\section{DATA AND ANALYSIS}
\label{s:dat}
Abundances of 15 elements have been derived from high-resolution, 
high-signal-to-noise ratio (SNR) spectroscopy of \cygab\ obtained with the 10-m 
Keck I telescope and HIRES echelle spectrograph (UT 1994 July 30).  The spectra 
are characterized by a nominal resolution of $R=\lambda / \Delta \lambda=45,000$ 
and SNR at the continuum near $\lambda 6700$ of 750 and 1050 for \cyga\ and \cygb, 
respectively.  A solar spectrum (Moon) was also obtained and has a SNR of 1500 
near $\lambda 6700$.  The data are the same as those used by 
\citet{1997AJ....113.1871K}, in which the observations, calibration scheme, and 
data reduction are fully described.

An updated version of the LTE spectral analysis package MOOG 
\citep{1973ApJ...184..839S} was used for the abundance analysis.  All 
abundances are derived from equivalent width (EW) measurements of atomic lines 
and the measurements were made using the one-dimensional spectrum analysis 
package SPECTRE \citep{1987BAAS...19.1129F}.  Carbon abundances are also derived 
by using the synthesis method to fit the observed spectra of two features 
($\lambda5086$ and $\lambda5135.6$) of the C$_2$ Swan system.  Stellar 
parameters were derived using excitation and ionization balance of \ion{Fe}{1} 
and \ion{Fe}{2} lines in the usual manner.

Our abundance and error analyses follow exactly those described in 
\citet{2011ApJ...732...55S}, where a more detailed description of the 
procedures can be found.  Final abundances-- given relative to solar abundances 
derived from our solar spectrum-- stellar parameters, and uncertainties 
for \cygab\ are given in Table \ref{tab:params}.  The adopted line list, 
equivalent width measures, and line-by-line abundances of each element for the 
Sun, \cyga, and \cygb\ are provided in Table \ref{tab:linelist}.

\section{RESULTS \& DISCUSSION}
\label{s:RandD}
The stellar parameters shown in Table \ref{tab:params}, we find \cygab\ to 
be physically similar, with A being slightly hotter and having a slightly lower 
surface gravity than B. The differences in parameters are $\Delta$\teff\ $=+43 \pm
45$ K, $\Delta \log g=-0.02 \pm 0.17$ dex, and $\Delta \xi=+0.10 \pm 0.11$ 
km s$^{-1}$.  While the parameters are the same within the uncertainties,
previous studies find consistently that \cyga\ is slightly hotter and has a lower
surface gravity than \cygb, suggesting that the small parameter differences are 
real.

\begin{figure}
\epsscale{0.9}
\plotone{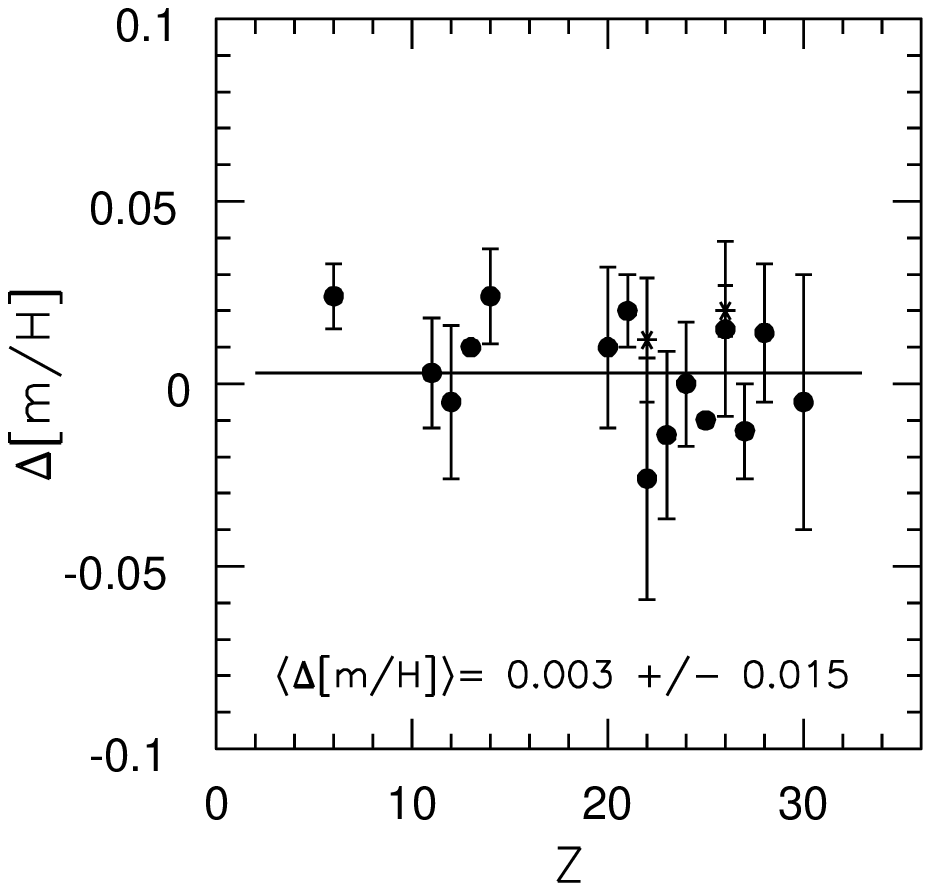}
\caption{\label{fig:dif} Abundance differences between \cyga\ and \cygb\ plotted
against atomic number ($Z$).  The six-pointed stars represent the abundances of
\ion{Ti}{2} and \ion{Fe}{2}.  The abundance difference for each element is 
the mean of the line-by-line abundance differences and is thus independent of 
the solar abundances; error bars are the standard deviations of the means.  The 
solid line is drawn at $\Delta$[m/H] $=0.003$, the mean abundance difference of 
all elements.}
\end{figure}

The [Fe/H] abundances are found to be indistinguishable within uncertainties,
with $\Delta$[Fe/H] $=+0.018 \pm 0.025$ ($\sigma$) dex, in agreement with previous
studies.  The difference in the Fe abundance, $\Delta$[Fe/H], is the average of
the line-by-line abundance differences of the \ion{Fe}{1} and \ion{Fe}{2} lines 
(difference of each individual line), as opposed to the difference in the mean 
abundances.  \citet{2001ApJ...553..405L} carried out a differential Fe abundance 
analysis of \cygab\  and found A to be enhanced in Fe relative to B by $0.025 
\pm 0.009$ dex.  However, \citet{2005PASJ...57...83T} conducted a similar 
differential analysis and found the metallicities to be identical at a level of 
$\lesssim 0.01$ dex.  Takeda also pointed out a possible systematic error in the 
analysis of \citet{2001ApJ...553..405L} that could account for the different 
results.  Abundances of the remaining elements derived here are also found to be 
indistinguishable, as seen in Table \ref{tab:params} and shown graphically in 
Figure \ref{fig:dif}.  The abundance differences shown in Figure \ref{fig:dif} 
are the means of the line-by-line differences for each element.  The mean 
abundance difference of all elements is $+0.003 \pm 0.015$ ($\sigma$) dex, with no 
element abundance differing by more than 0.026 dex between the two stars.

Given the marked agreement in the abundances of \cygab\ for the 15 elements 
studied here, it seems likely that these two binary components are chemically
identical save the factor of $\geq 4.5$ difference in their Li abundances 
\citep{1997AJ....113.1871K}.  The chemical homogeneity suggests that the Li 
abundance difference is not primordial but rather due to some physical process
during the lifetime of the system.  \citet{2001ApJ...553..405L} suggested that
accretion of planetary material by A could explain its enhanced Li
abundance relative to B.  \citet{2010A&A...521A..44B} have alternatively 
demonstrated that episodic accretion onto a young star can affect its internal 
structure and increase its core temperature, resulting in enhanced surface Li 
depletion.  The similar chemical compositions of \cygab\ argues against any 
differential accretion onto either of the stars having occurred.

The disparate Li abundances of \cygab\ are more likely the result of 
rotationally-induced mixing and differences in angular momentum evolution.  
\citet{1997AJ....113.1871K} argue that non-standard slow mixing on the main 
sequence, possibly related to rotation, can account for the stars' low absolute 
Li abundances.  The difference in the Li abundances of \cygab\ would then be due 
to differences in $J_{\mathrm{o}}$ and/or the rates of angular momentum loss.  
\citet{1997AJ....113.1871K} suggest that planet formation could affect the angular 
momentum evolution of the host star.  Recent modeling efforts do indeed 
demonstrate the plausibility of this assertion 
\citep[e.g.,][]{2008A&A...489L..53B,2010A&A...519L...2E}.  For instance, 
\citet{2008A&A...489L..53B} shows that shear-induced turbulence due to 
core-envelope decoupling can result in enhanced Li depletion in solar-type stars 
and that stars with slow rotation rates on the zero-age main sequence (ZAMS) 
have longer core-envelope coupling timescales than fast rotators.  Slow rotators 
are thus expected to deplete more Li than fast rotators.  Bouvier further
demonstrates that, compared to stars with short-lived circumstellar disks, stars 
with longer-lived disks will experience more angular momentum loss via magnetic 
star-disk interactions and will arrive on the ZAMS as more slowly rotating stars 
and thus have lower Li abundances.

This could explain, at least qualitatively, why two otherwise physically similar
and chemically homogeneous stars such as \cygab\ could have significantly 
different Li abundances.  Whereas the presence of a massive planet orbiting 
\cygb\ evidently requires a disk with a lifetime sufficient to form such a 
planet, the lack of a detected planet orbiting \cyga\ suggests that, if this 
star had a disk, its physical properties were such that planet formation was 
inhibited.  Both observational \citep[e.g.,][]{1996ApJ...458..312J} and 
computational \citep[e.g.,][]{2005MNRAS.363..641M} studies suggest disk 
structure and as a result planet formation are disrupted in binary systems with 
separations less than 100 AU.  If the disk of \cyga\ was truncated by its M
dwarf companion, determined to be at $\sim 70$ AU, its shorter lifetime compared 
to the planet-forming disk of \cygb\ may have resulted in less Li destruction.  
While the lower Li abundance of \cygb\ relative to \cyga\ is consistent with 
this scenario, results of observational studies aimed at tying enhanced Li 
depletion to the presence of planets have not reached a consensus on the matter 
\citep[e.g.,][]{2009Natur.462..189I,2010ApJ...724..154G,2010A&A...519A..87B}. 
Nonetheless, the case of \cygab\ is intriguing as it may be an ideal system 
for further studies of the possible connection between binarity, planet 
formation, and Li depletion.

\subsection{Abundance Trends with Condensation Temperature of the Elements}
\label{ss:tc}
The fact that no planet has heretofore been discovered around \cyga\ does not 
preclude the existence of a planet orbiting this star.  However, the chemical 
composition of \cygab\ may place additional constraints on the existence of such a 
planet.  \citet{2009ApJ...704L..66M} have demonstrated that the Sun is deficient 
in refractory elements relative to volatile elements compared to a sample of solar 
twins.  Moreover, the deficiencies are correlated with the condensation 
temperature of the elements (\tc) such that the abundances of refractory elements 
(\tc$\gtrsim 900$ K) decrease with increasing \tc.  \citet{2009ApJ...704L..66M} 
suggest that the abundance pattern is due to dust condensation and terrestrial 
planet formation in the proto-solar nebula.  Follow-up studies 
\citep{2009A&A...508L..17R,2010A&A...521A..33R} including larger samples of solar 
twins and analogs found that the abundance patterns of $\sim85$\% of the stars 
analyzed differ from the Sun, i.e., they have increasing abundances of refractory 
elements as a function of \tc.  The authors speculate that the remaining 
$\sim 15$\% with flat or decreasing trends are potential terrestrial planet hosts.

We have recently extended the analysis of abundances versus \tc\ trends to a
sample of 10 stars known to host giant planets \citep{2011ApJ...732...55S}.  The 
slopes of linear least-squares fits to the [m/H]-\tc\ trends were compared to 
similar slopes for a sample of 121 stars with and without known giant planets 
from \citet{2010MNRAS.407..314G}; the distribution of slopes as a function of 
[Fe/H] for this larger sample was taken as the general trend arising from Galactic 
chemical evolution.  Four of the 10 stars in our sample have very close-in giant 
planets (three at 0.05 AU) and are found to have positive slopes that fall above 
the general trend defined by the Gonzalez et al. data.  These stars are speculated 
to have accreted refractory-rich planet material sometime during the evolution of 
their planetary systems.  Abundance trends with \tc\ then may not only indicate 
the presence of terrestrial planets but also provide clues to the architecture of 
a planetary system and/or evolution thereof.  The remaining six stars from 
\citet{2011ApJ...732...55S} have negative slopes, possibly indicating the 
presence of terrestrial planets, but the slopes fall along the general trend of 
Galactic chemical evolution and thus may not be related to planet formation.

\begin{figure}
\epsscale{0.9}
\plotone{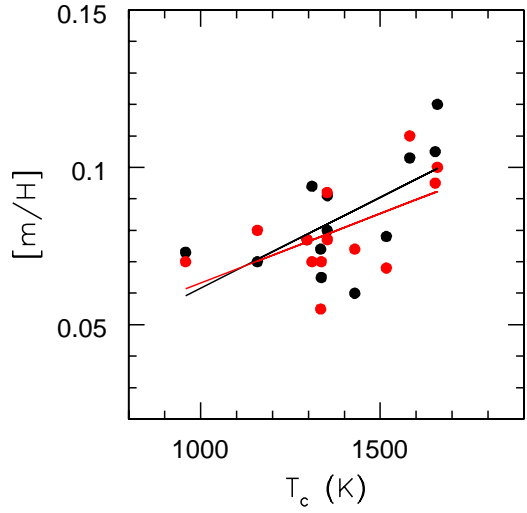}
\caption{\label{fig:tc}Relative abundances as a function of condensation 
temperature of the elements.  Abundances of \cygab\ are plotted as black and red
points, respectively.  The solid lines are linear least-squares fits to the data 
and have positive slopes that are indistinguishable: $m_{\mathrm{A}}=5.77 \pm 
2.08 \times 10^{-5}$ dex K$^{-1}$ and $m_{\mathrm{B}}=4.42 \pm 1.94 \times 
10^{-5}$ dex K$^{-1}$ for A (black) and B (red), respectively.}
\end{figure}

The abundances of \cygab\ are plotted versus \tc\ in Figure \ref{fig:tc}.  Only
the refractory elements (\tc$\gtrsim 900$ K) are considered, because it is among 
these elements that the putative planet signature has been detected 
\citep{2009ApJ...704L..66M}.  The abundances are plotted against 50\% \tc\ from 
\citet{2003ApJ...591.1220L}.  Slopes of linear least-squares fits are positive and 
identical within the uncertainties:
$m_{\mathrm{A}}=5.77 \pm 2.08 \times 10^{-5}$ dex K$^{-1}$ and 
$m_{\mathrm{B}}=4.42 \pm 1.94 \times 10^{-5}$ dex K$^{-1}$ for \cyga\ and
\cygb, respectively.

Positive slopes in the [m/H]-\tc\ relations for \cygab, in the interpretation 
of \citet{2009A&A...508L..17R}, imply that these solar twins are not terrestrial 
planet hosts.  Continued RV monitoring have failed to yield additional planet 
signatures for either \cygab, but the sensitivity of the ground-based RV 
observations may not be sufficient to detect small terrestrial planets.  
\citet{2007AJ....134.1276W} investigated the likelihood that additional planets 
could survive in the \cygb\ system given the large eccentricity of \cygb\ b.  
Using test-particle simulations, they found that particles only remained in stable 
orbits inside 0.3 AU, leaving open the possibility that short period planets may 
exist in this system.  However, combining the numerical simulations with RV 
monitoring data, planets with masses $M \sin i \gtrsim 2$ Neptune mass with 
periods of less than about 100 days (roughly corresponding to $a=0.3$ AU) can be 
excluded at the 99\% confidence level.

The physical process(es) responsible for the large eccentricities characteristic
of many of the known extrasolar planets, including \cygb\ b, is currently not 
well constrained.  Planet-disk interactions have been investigated, but 
simulations generally result in the dampening of orbital eccentricities and do 
not reproduce the observed planet eccentricity distribution 
\citep[e.g.,][]{2010A&A...523A..30B}.  An alternative explanation is dynamical 
instabilities resulting from planet-planet scattering.  Simulations of
multi-planet systems can produce planets with highly eccentric orbits, and more 
importantly, they can reproduce the observed extrasolar planet eccentricity 
distribution \citep[e.g.,][]{2008ApJ...686..621F,2009ApJ...699L..88R}.  For 
\cygb\ b, \cyga\ may be the culprit.  Secular interactions with a distant 
stellar companion have been shown to produce long-period oscillations in the 
eccentricities of a planet orbiting the companion binary star 
\citep[the so called Kozai mechanism;][]{2005ApJ...627.1001T}.
\citet{1997Natur.386..254H} and \citet{1997ApJ...477L.103M} have independently
demonstrated that such a mechanism is plausibly responsible for the large
eccentricity of \cygb\ b.

A possible consequence of induced eccentricity enhancement is the ejection of 
disk or planet material in the inner region of the system, disrupting terrestrial 
planet formation.  Simulations testing the effects of giant planets with eccentric 
orbits on the formation of terrestrial planets generally show a near complete 
clearing out of inner planetary material and thus no terrestrial planet formation 
\citep[e.g.,][]{2005ApJ...620L.111V,2011A&A...530A..62R}.  In particular, 
\citet{2011A&A...530A..62R} reported that in simulations in which a giant planet 
scattered to a minimum periastron distance of $< 1.3$ AU, all of the terrestrial 
material in those systems was destroyed.  Extending this result to \cygb\ b, the 
periastron of which is $r_{\mathrm{p}}=0.52$ AU based on the most recently derived 
orbital parameters \citep[$e=0.689$ and $a=1.68$ AU;][]{2007ApJ...654..625W}, no 
terrestrial planet material would be expected to have survived around \cygb.  This 
is consistent with the implication of the positive slopes in the [m/H]-\tc\ 
relations for \cygab.

\section{Conclusions}
\label{s:con}
We have presented the results of a detailed abundance analysis of the solar twins
\cygab, the second of which is host to a giant planet.  Aside from a factor of 
$\sim 4.5$ difference in Li abundances, the two stars are found to be otherwise 
chemically identical based on the 15 elements considered.  Slopes in the 
[m/H]-\tc\ relations are also statistically identical and are another indication 
that \cygab\ are chemically homogeneous.  The stark consistency of the 
compositions of these stars suggest that the physical process(es) responsible for 
the enhanced Li depletion in B did not alter the abundances of other elements.  
This argues against any kind of accretion related mechanism and supports 
differences in internal mixing efficiencies possibly related to different angular 
momentum evolutions as the most likely explanation for the disparate Li 
abundances.  Enhanced Li depletion in B can be plausibly tied to the presence of 
its giant planet, as predicted by rotational stellar evolution models; however, 
the mixed observational results regarding Li abundances of planet host stars cloud 
this issue.  More work is clearly required to understand how star-disk 
interactions and/or planet formation does or does not increase Li depletion in 
planet host stars.

The chemical homogeneity of \cygab, combined with the heretofore lack of 
detected planets around \cyga, further suggests that the planet formation 
process did not affect the bulk composition of \cygb.  Since the discovery 
that stars with giant planets tend to be more metal-rich than stars without known 
planets 
\citep{1997MNRAS.285..403G,1998A&A...334..221G,2004A&A...415.1153S,2005ApJ...622.1102F}, 
countless abundance studies of host stars have aimed to identify possible chemical 
vestiges of the planet formation process.  As described above, Li may be one of 
these.  As for the overall metallicity of planet hosts, the result for \cygab\ 
adds to the considerable evidence indicating that the planet-metallicity 
correlation for stars with giant planets is intrinsic in nature and does not 
arise from processes, such as accretion of solid-body material, associated with 
the formation and evolution of giant planets.  Furthermore, it appears that the 
abundances of individual elements heavier than Li \citep[with the possible 
exception of Be and B, the abundances of which can also be depleted by internal 
mixing mechanisms, depending on the depth and efficiency of the 
mixing;][]{1998ApJ...498L.147D,2005ApJ...621..991B} are also not affected by 
planet formation, at least in systems like \cygb.

The physical characteristics of 16 Cygni make it an ideal system to test and 
constrain planet formation models.  Most tellingly, the conditions necessary for 
planet formation apparently were present for \cygb\ but not \cyga, despite their 
physical and chemical similarities.  We have discussed empirical and 
computational results that can possibly account for the observed characteristics 
of the system, including the lack of a detected planet around \cyga, the 
enhanced Li depletion of \cygb, and the eccentricity of the planet \cygb\ b, and 
that imply that neither \cyga\ nor \cygb\ is a terrestrial planet host.  Future 
efforts that can combine all of these attributes into a single model will 
represent a significant achievement in understanding the formation and evolution 
of planetary systems.

\acknowledgements
S.C.S. acknowledges support provided by the NOAO Leo Goldberg Fellowship; NOAO
is operated by AURA, Inc. under a cooperative agreement with the NSF.  L.G. 
acknowledges support by the PAPDRJ - CAPES/FAPERJ Fellowship.  J.R.K. acknowledges 
support by NSF award AST-0908342.  

{\it Facilities:} \facility{KECK(HIRES)}

\newpage

\begin{deluxetable}{lrrrr}
\tablecolumns{5}
\tablewidth{0pt}
\tablecaption{Stellar Parameters \& Abundances\label{tab:params}}
\tablehead{
	\colhead{Parameter\tablenotemark{a}}&
	\colhead{}&
	\colhead{\cyga}&
	\colhead{}&
	\colhead{\cygb}
	}
\startdata
$T_{\mathrm{eff}}$ (K) && 5796 $\pm 34$          && 5753 $\pm 30$   \\
$\log g$ (cgs)         && 4.38 $\pm 0.12$        && 4.40 $\pm 0.12$ \\
$\xi$ (km s$^{-1}$)    && 1.45 $\pm 0.07$        && 1.35 $\pm 0.08$ \\
$[$Fe/H$]$  \dotfill   &&+0.07 $\pm0.01$\tablenotemark{b} $\pm 0.05$\tablenotemark{c} &&+0.05 $\pm0.01 \, \pm 0.05$ \\[0.15truein]

$[$C/H$]$   \dotfill   &&+0.10 $\pm0.03 \, \pm 0.05$ &&+0.08 $\pm0.03 \, \pm 0.05$ \\
$[$Na/H$]$  \dotfill   &&+0.07 $\pm0.00 \, \pm 0.03$ &&+0.07 $\pm0.00 \, \pm 0.03$ \\
$[$Mg/H$]$  \dotfill   &&+0.07 $\pm0.04 \, \pm 0.05$ &&+0.07 $\pm0.04 \, \pm 0.03$ \\
$[$Al/H$]$  \dotfill   &&+0.11 $\pm0.02 \, \pm 0.03$ &&+0.10 $\pm0.02 \, \pm 0.03$ \\
$[$Si/H$]$  \dotfill   &&+0.09 $\pm0.01 \, \pm 0.01$ &&+0.07 $\pm0.01 \, \pm 0.01$ \\
$[$Ca/H$]$  \dotfill   &&+0.08 $\pm0.01 \, \pm 0.04$ &&+0.07 $\pm0.01 \, \pm 0.04$ \\
$[$Sc/H$]$  \dotfill   &&+0.12 $\pm0.01 \, \pm 0.07$ &&+0.10 $\pm0.01 \, \pm 0.07$ \\
$[$Ti/H$]$  \dotfill   &&+0.10 $\pm0.01 \, \pm 0.07$ &&+0.11 $\pm0.01 \, \pm 0.07$ \\
$[$V/H$]$   \dotfill   &&+0.06 $\pm0.02 \, \pm 0.04$ &&+0.07 $\pm0.02 \, \pm 0.04$ \\
$[$Cr/H$]$  \dotfill   &&+0.08 $\pm0.02 \, \pm 0.04$ &&+0.08 $\pm0.02 \, \pm 0.03$ \\
$[$Mn/H$]$  \dotfill   &&+0.07 $\pm0.03 \, \pm 0.04$ &&+0.08 $\pm0.03 \, \pm 0.04$ \\
$[$Co/H$]$  \dotfill   &&+0.08 $\pm0.02 \, \pm 0.04$ &&+0.09 $\pm0.02 \, \pm 0.03$ \\
$[$Ni/H$]$  \dotfill   &&+0.09 $\pm0.01 \, \pm 0.02$ &&+0.08 $\pm0.01 \, \pm 0.02$ \\
$[$Zn/H$]$  \dotfill   &&+0.10 $\pm0.02 \, \pm 0.04$ &&+0.10 $\pm0.02 \, \pm 0.03$ \\

\enddata

\tablenotetext{a}{Adopted solar parameters: \teff\ $=5777$ K, $\log g=4.44$, 
and $\xi=1.38$ km s$^{-1}$.}
\tablenotetext{b}{$\sigma_{\mu}$}
\tablenotetext{c}{$\sigma_{Total}$-- quadratic sum of $\sigma_{\mu}$ and uncertainties due to uncertainties in \teff, $\log g$, and $\xi$.}

\end{deluxetable}

\begin{deluxetable}{lcccccccrccrccrc}
\tablecolumns{16}
\tablewidth{0pt}
\rotate
\tablecaption{Lines Measured, Equivalent Widths, and Abundances\label{tab:linelist}}
\tablehead{
     \colhead{}&
     \colhead{}&
     \colhead{$\lambda$}&
     \colhead{}&
     \colhead{$\chi$}&
     \colhead{}&
     \colhead{}&
     \colhead{}&
     \colhead{}&
     \colhead{}&
     \colhead{}&
     \multicolumn{2}{c}{\cyga}&
     \colhead{}&
     \multicolumn{2}{c}{\cygb}\\
     \cline{12-13} \cline{15-16}\\
     \colhead{Ion}&
     \colhead{}&
     \colhead{(\AA)}&
     \colhead{}&
     \colhead{(eV)}&
     \colhead{}&
     \colhead{$\log \mathrm{gf}$}&
     \colhead{}&
     \colhead{EW$_{\odot}$}&
     \colhead{$\log N_{\odot}$}&
     \colhead{}&
     \colhead{EW}&
     \colhead{$\log N$}&
     \colhead{}&
     \colhead{EW}&
     \colhead{$\log N$}
     }
\startdata
\ion{C}{1}  && 5052.17 && 7.68 && -1.304 &&  31.8 & 8.43 &&  39.6 & 8.56 &&  37.0 & 8.54 \\
            && 5380.34 && 7.68 && -1.615 &&  19.8 & 8.46 &&  25.0 & 8.58 &&  22.7 & 8.55 \\
            && 6587.61 && 8.54 && -1.021 &&  13.5 & 8.41 &&  18.5 & 8.57 &&  16.7 & 8.54 \\
\ion{Na}{1} && 5682.63 && 2.10 && -0.700 && 105.0 & 6.21 && 109.7 & 6.29 && 110.2 & 6.27 \\
            && 6154.23 && 2.10 && -1.560 &&  38.4 & 6.29 &&  42.7 & 6.36 &&  44.5 & 6.37 \\
            && 6160.75 && 2.10 && -1.260 &&  58.1 & 6.26 &&  62.9 & 6.33 &&  64.2 & 6.33 \\
\ion{Mg}{1} && 4730.03 && 4.35 && -2.523 &&  74.1 & 7.91 &&  80.0 & 8.00 &&  80.3 & 7.99 \\
            && 5711.09 && 4.35 && -1.833 && 104.3 & 7.60 && 106.7 & 7.64 && 108.6 & 7.66 \\

\enddata

\tablecomments{Table 2 is published in its entirety in the electronic edition of
The Astrophysical Journal Letters.  A portion is shown here for guidance regarding 
its form and content.}

\end{deluxetable}

\end{document}